\documentclass[conference]{IEEEtran}
\usepackage{upgreek}

\usepackage{url}

\usepackage{amsmath}

\usepackage[T1]{fontenc}

\usepackage{array}

\usepackage{graphicx}

\usepackage{subfig}
\usepackage{multirow}
\usepackage{array}
\usepackage{amssymb}

\usepackage[utf8]{inputenc}

\usepackage{url}
\graphicspath{{./pdf/}{./jpeg/}{./images/}}
   \DeclareGraphicsExtensions{.pdf,.jpeg,.jpg,.png}

\ifCLASSINFOpdf
\else
\fi
\begin{document}
%
\title{An Analysis of BitTorrent Cross-Swarm Peer Participation and Geolocational Distribution}

\author{\IEEEauthorblockN{Mark Scanlon}
\IEEEauthorblockA{School of Computer Science \& Informatics,\\
University College Dublin,\\
Belfield, Dublin 4, Ireland.\\
Email: mark.scanlon@ucd.ie}
\and
\IEEEauthorblockN{Huijie Shen}
\IEEEauthorblockA{Software School,\\
Fudan University, 
\\Shanghai, China.\\
Email: 10302010068@fudan.edu.cn}}


%


\maketitle

\begin{abstract}
Peer-to-Peer (P2P) file-sharing is becoming increasingly popular in recent years. In 2012, it was reported that P2P traffic consumed over 5,374 petabytes per month, which accounted for approximately 20.5\% of consumer internet traffic. TV is the popular content type on The Pirate Bay (the world's largest BitTorrent indexing website). In this paper, an analysis of the swarms of the most popular pirated TV shows is conducted. The purpose of this data gathering exercise is to enumerate the peer distribution at different geolocational levels, to measure the temporal trend of the swarm and to discover the amount of cross-swarm peer participation. Snapshots containing peer related information involved in the unauthorised distribution of this content were collected at a high frequency resulting in a more accurate landscape of the total involvement. The volume of data collected throughout the monitoring of the network exceeded 2 terabytes. The presented analysis and the results presented can aid in network usage prediction, bandwidth provisioning and future network design.
\end{abstract}


%
\IEEEpeerreviewmaketitle

\section{Introduction}
In recent years, P2P file-sharing has become very popular around the world. Users choosing to partake in P2P file-sharing can download and share various digital content, such as videos, music, software and books from other users in the network. According to Cisco's estimation \cite{cisco}, P2P file transfer consumed 5,374 petabytes per month in 2012, which is about 20.5\% of consumer internet traffic. Cisco predicts that the pattern of growth of P2P traffic will continue for the foreseeable future. Due to the increasing volume of streaming video, the proportion of P2P file-sharing traffic in comparison with global traffic usage will decrease. BitTorrent is the one of the most popular P2P protocols used today. It is reported to be responsible for over 53\% of file-sharing's total bandwidth and over 3\% of total bandwidth worldwide by August 2013 \cite{paloalto}. BitTorrent, Inc. announced that the company's flagship BitTorrent mainline and $\mu$Torrent software clients have grown to over 150 million monthly active users worldwide in January 2012 \cite{bt}.

The BitTorrent protocol \cite{btspec} is a P2P protocol for efficient distribution and replication of files, and is advantageous in the distribution of large files. Traditional sources for file distribution allow users to download directly from the source's server. As more downloading requests are received from users, the server's bandwidth usage rises dramatically. This in turn significantly increases the traffic costs for the owner. As an alternative, BitTorrent facilitates users in the simultaneous downloading and uploading of files with each other. Each BitTorrent client serves as both a client and a server. Therefore, the traffic required of the publisher's machine stays relatively low when a large number of users are simultaneously downloading a particular file. 

\subsection{Contribution of this Work}
\label{contribution}
The popularity of P2P file-sharing and the BitTorrent protocol has captured the attention and interest of many researchers. Many studies have been performed on the characteristics of IP participation and pattern of BitTorrent's geolocational distribution \cite{scanlon,chen,salvador} and flashcrowds \cite{zhang,pouwelse,guo}. The contribution of this work outlined as part of this paper complements and expands upon this body of related work in two main ways: 
\begin{enumerate}
\item Crawl Frequency -- Swarm information is collected at much shorter intervals when compared to previous work. This helps in improving the accuracy of the results achieved. For the purposes of enumeration and cross-swarm analysis, more frequent crawls result in most accurate results.
\item Significant Cross-swarm Analysis -- Through the design and utilisation of a NoSQL database, queries on the gathered data which were previously computationally prohibitive are now possible. As a result, a number of new cross-swarm metrics were discovered.
\end{enumerate}

The results from the analysis of this P2P traffic could be beneficial in a number of areas:
\begin{enumerate}
\item The prediction of network utilisation and the resultant provisioning of bandwidth across local access hubs increasing the Quality of Service (QoS) provided by Internet Service Providers.
\item Gaining a better understanding of P2P network traffic patterns can aid in the design of more robust, performant P2P networks and P2P client application implementations.
\end{enumerate}

\section{The BitTorrent Protocol}
In order to understand how BitTorrent operates, a number of terminologies must first be defined:
\begin{enumerate}
\item Peer: An active BitTorrent client that other peers can connect with to share content.
\item Leecher: A peer that is downloading a specific piece of content.
\item Seeder: A peer that has finished downloading a certain file and is currently uploading parts of the file to leechers.
\item Swarm: All peers sharing the same file form a swarm, incorporating both seeders and leechers.
\item Tracker: A server responsible for keeping track of the peers in a swarm. Each peer communicates with its tracker periodically about how much it has uploaded/downloaded, etc. The tracker will respond with the number of seeders and leechers in the swarm, as well as some specific information about some active peers in the swarm, etc.
\item \texttt{.torrent} file: A meta-data file encoded using a bespoke dictionary stored using ``bencoding''. This file describes the basic properties of the content, such as name, length, piece length, etc. It also contains the announce URL of the tracker. BitTorrent divides a file into small parts of fixed size (512kb by default) called pieces, whose SHA-1 checksums are specified in the \texttt{.torrent} file.
\end{enumerate}


As BitTorrent is a P2P protocol, users need to install a BitTorrent client on their computer to download files that are distributed using BitTorrent protocol. BitTorrent and $\mu$Torrent are two commonly used clients, both of which are developed by BitTorrent, Inc. To differentiate between the protocol and the application and for the purpose of this paper, ``BitTorrent'' will refer solely to the BitTorrent protocol. Once a client is installed, users download a \texttt{.torrent} file of the desired content from a BitTorrent indexing website and open the file using a BitTorrent client to download the content. The Pirate Bay is the largest public BitTorrent indexing website with 5,190,408 torrents and 37,947,184 peers in August 2013 \cite{tpb}. It has an enormous collection of torrents of all kinds of digital content, including TV shows, movies, music, games, software, books, etc. In the spring TV show season of 2013, there are around 5,200,000 downloads per episode of Game of Thrones while the number of viewers in US is 5,500,000 \cite{torrentfreak}. In order to identify the geographic distribution of these downloads and the temporal swarm size trends, a month long data gathering on three popular TV shows was conducted, as outlined in Section \ref{investigationdesign}.

\section{Data Gathering Process}
\label{investigationdesign}

TV shows and movies are the most popular content on The Pirate Bay \cite{tpb}. Unlike movies, a TV show generally consists of several seasons and several episodes for each season. Therefore, the cumulative downloads of a TV show can be much more than that of a movie. The data gathered focuses on popular TV shows, which attract a significant number of users and subsequently represent a large proportion of BitTorrent's total global bandwidth.

The characteristics decided upon for the content selection to be monitored were:
\begin{itemize}
\item A large number of downloads for previous episodes.
\item A definite schedule for the following episodes.
\item At least 2 episodes to broadcast after the data gathering process has begun.
\end{itemize}

\begin{table}[!t]
\begin{footnotesize}
\renewcommand{\arraystretch}{1.0}
\caption{Torrent Information}
\label{torrentinformation}
\makebox[\linewidth]{
\begin{tabular}{|c|c|c|c|}
\hline
\textbf{Name} & \textbf{Size} 
\\
\hline
Breaking.Bad.S05E09.HDTV.x264-ASAP.mp4 & 327.93 MB 
\\
\hline
Breaking.Bad.S05E09.720p.HDTV.x264-IMMERSE[rarbg] & 1.09 GB 
\\
\hline
Breaking.Bad.S05E10.HDTV.x264-ASAP.mp4 & 304.23 MB 
\\
\hline
Breaking.Bad.S05E10.720p.HDTV.x264-IMMERSE.mkv & 1.07 GB 
\\
\hline
Breaking Bad S05E11 HDTV x264-ASAP[ettv] & 318.09 MB 
\\
\hline
Breaking.Bad.S05E11.720p.HDTV.x264-IMMERSE.mkv & 1.14 GB 
\\
\hline
Dexter S08E07 HDTV x264-ASAP[ettv] & 335.51 MB 
\\
\hline
Dexter.S08E07.720p.HDTV.x264-IMMERSE.mkv & 1.21 GB 
\\
\hline
Dexter S08E08 HDTV x264-ASAP[ettv] & 316.28 MB 
\\
\hline
Dexter.S08E08.HDTV.x264-ASAP.mp4 & 316.28 MB 
\\
\hline
Dexter S08E09 HDTV x264-ASAP[ettv] & 350.47 MB 
\\
\hline
Dexter.S08E09.HDTV.x264-ASAP.mp4 & 350.47 MB
 \\
\hline
True Blood S06E09 Life Matters WEB DL XviD-FUM[ettv] & 435.65 MB 
\\
\hline
True.Blood.S06E09.HDTV.x264-EVOLVE.mp4 & 623.11 MB 
\\
\hline
True Blood S06E10 Radioactive WEB-DL XviD-FUM[ettv] & 424.18 MB 
\\
\hline
True.Blood.S06E10.HDTV.x264-KILLERS.mp4 & 506.98 MB 
\\
\hline
\end{tabular}
}
\end{footnotesize}
\end{table}

Based on the standard outlined above, three TV shows were selected: Breaking Bad (Season 5), Dexter (Season 8) and True Blood (Season 6). For each episode, there are usually over 20 different torrents available created by different uploaders/release groups. Due to the large volume of data collected for each swarm, the two most popular torrents were selected for each episode, i.e., two distinct releases of the same episode. At noon (GMT) on the day of the episode release, the two largest swarms were easily identifiable and added to the monitoring crawler. From subsequent observations, these torrents maintained their high popularity throughout the data gathering window. Table~\ref{torrentinformation} gives a full list of selected torrents, including names (more accurately, the value of the \texttt{save\_as} field in the \texttt{.torrent} files), sizes and uploading time shown on The Pirate Bay.

\subsection{Crawling}
Peer information is gathered during a crawl for each piece of content and stored in XML files. These XML files are parsed and the extracted peer information is stored in results database for subsequent analysis. The outline of the crawling application is beyond the scope of this paper, but the process consists of the following four steps:
\begin{enumerate}
\item Connect to The Pirate Bay and acquire the torrent specific information required for processing on the day of release. Since The Pirate Bay no longer provides direct \texttt{*.torrent} downloading from March 2012, the magnet URI is used (this contains most of the same information contained in a \texttt{*.torrent} file).
\item Connect to each swarm sequentially and identify each of the IP addresses currently active in the swarm until
no new IPs are found.
\item Store the IPs and related information in a XML file for each torrent.
\item On average after two minutes, the entire process was restarted again at Step 2. This two minute window was the average time required to crawl each of the torrents investigated.
\end{enumerate}

\section{Database Design}
\begin{figure}[!b]
\centering
\includegraphics[width=3in]{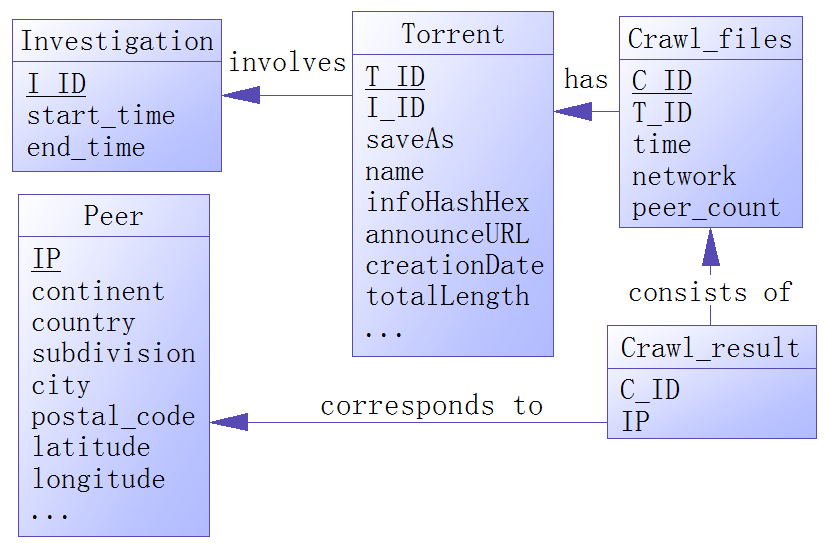}
\caption{The Data Model in Relational Database}
\label{DB}
\end{figure}
This data analysis expands upon previous similar work by Scanlon and Kechadi \cite{scanlon}, which adopted MySQL as the database management system. As part of the analysis of the results for that work, the database grew to such a large size that the insertion and query performance were dramatically impacted. Since the analysis presented in this paper required a much larger data volume than gathered previously, relational databases were deemed incapable of the necessary performance. As a result, a NoSQL database was selected, MongoDB.

With the popularity of social media and the recognition of issues surrounding ``big data'', traditional relational databases face a big challenge to store and process huge amounts of data. As an alternative, various NoSQL databases are evaluated. Compared with SQL Databases, NoSQL databases are schemaless and they maintain high performance in concurrent read/write even if data volume is huge. For different applications there are different suitable types of NoSQL databases, including key-value store, document store, graph store and column store. The data processed for analysis in this paper consisted of IPs, related geolocational and temporal information. A document-oriented NoSQL database met the storage and performance requirements.

A large proportion of queries used were related with the number of distinct IPs on country, US state and city level of a particular torrent (or torrents) in one month. From the perspective of a relational database, the data model would be like the one in Figure~\ref{DB}. Hence, the simplest design is to break the `join' between \texttt{crawl\_result} and other tables and store all the data needed in \texttt{crawl\_result}, e.g., \texttt{IP}, \texttt{torrent\_id}, \texttt{country}, \texttt{state} (if it is a US/Canadian IP), \texttt{city}, \texttt{longitude}, \texttt{latitude} and \texttt{time}. 

However, MongoDB supports distinct operation on a small scale of distinct values. Testing showed that distinct queries which would return millions of results cannot be executed in a timely manner. Instead, 16 fields were added to \texttt{peer} (each IP address is stored only once in this collection): \\\texttt{t\_1}, \texttt{t\_2}, \texttt{...} , \texttt{t\_16}. This correspond to the IDs assigned to each torrent in the database. Field \texttt{t\_i} will be true in collection \texttt{peer} if the peer is detected active in the swarm of torrent \texttt{t\_i}. Otherwise, this field is set to false. In this way, the number of distinct IPs in single or multiple swarms can be easily calculated. The geolocational information stored in \texttt{peer} is derived from GeoIP2 \cite{maxmind}. GeoIP2 database offers geolocational information for a given IP. GeoIP, the former version, is 99.8\% accurate on a country level, 90\% accurate on a state level in the US, and 83\% accurate for cities in the US within a 40 kilometre radius and GeoIP2 improves slightly on these accuracy levels.

The statistics produced included the peer-level geolocational distribution and the temporal trend of the whole swarm size. The significance was also measured of how much peers from Australia, Europe and North America have on the swarms overall size. Based on the result of a small-scale test, these three regions play an important role in influencing the overall swarm size. The field \texttt{peer\_count} stored in the XML files greatly reduces the work to count distinct IPs for a given instant in \texttt{crawl\_result}. Similarly, \texttt{AUSCount}, \texttt{EuroCount} and \texttt{NACount} were also calculated. When a XML file is parsed, the country for each IP is get from GeoIP2 database, if the location is among the three continental categories, the corresponding value will be incremented. These three fields will be inserted into \texttt{crawl\_files} together with other information when parsing ends.

The design of collections \texttt{peer} and \texttt{crawl\_files} are outlined below:


\begin{enumerate}
\item \texttt{peer(IP, country, state, city, ISP, longitude, latitude, t\_1, ... , t\_16)}
\item \texttt{crawl\_files(time, network, peer\_count, torrent\_id, EuroCount, NACount, AUSCount)}
\end{enumerate}

\section{Results and Analysis}
The data gathering process began on 12 August 2013 and ended on 12 September 2013. During this month long window, 16 swarms were selected, as outlined in Table~\ref{torrentinformation}. In total 1,272,194,701 peer hits were gathered and stored in \texttt{crawl\_result} while the number of distinct IPs was 6,299,695. In other words, each IP was discovered in \texttt{crawl\_result} over 200 times on average (resulting in an average time of swarm activity of over 6.6 hours). 

Before the presentation of the results, some assumptions must be accepted for their interpretation. Firstly, although GeoIP2 is one of the most accurate databases for IP-to-geolocation conversion, the possibility still exists that it provides inaccurate geolocational information on a small number of the IP addresses detected. Secondly, a small proportion of the end users may have employed a proxy, dynamic IP allocation, or IP sharing during the data gathering process. Thus, the precise number of end users may be marginally different to the numbers reported below.

\subsection{IP Participation at the Swarm Level}

\begin{table}[!h]
\begin{footnotesize}
\renewcommand{\arraystretch}{1.0}
\caption{IP Participation at Swarm Level}
\label{aug12th}
\makebox[\linewidth]{
\begin{tabular}{|c|c|c|}
\hline
\textbf{Swarm name} & \textbf{Distinct IPs} & \textbf{Overall \%}\\
\hline
\shortstack{Breaking.Bad.S05E09.\\HDTV.x264-ASAP.mp4} & \shortstack{1,648,666} & 26.17\%\\
\hline
\shortstack{Breaking.Bad.S05E09.720p.\\HDTV.x264-IMMERSE[rarbg]} & 347,814 & 5.52\%\\
\hline
\shortstack{Dexter S08E07 HDTV \\x264-ASAP[ettv]} & 983,860 & 15.62\%\\
\hline
\shortstack{Dexter.S08E07.720p.\\HDTV.x264-IMMERSE.mkv} & 311,144 & 4.94\%\\
\hline
\shortstack{True.Blood.S06E09.\\HDTV.x264-EVOLVE.mp4} & 903,936 & 14.35\%\\
\hline
\shortstack{True Blood S06E09 Life Matters\\ WEB DL XviD-FUM[ettv]} & 206,774 &3.28\%\\
\hline
\end{tabular}
}
\end{footnotesize}
\end{table}

For each TV series, the two torrents downloaded on August 12th 2013 were tracked for one month. Table~\ref{aug12th} lists the number of distinct active IPs in these swarms. As can be seen, even the biggest 2 swarms for each episode can vary a lot in swarm size. The largest swarm attracts most of the peers interested in each episode. In the case of Breaking Bad and Dexter, torrents of relatively smaller file are more popular than that of larger ones - with over 2 times more distinct IPs. However, such a situation is quite different as to True Blood. A reasonable explanation is that the file related with torrent \texttt{True Blood S06E09 Life Matters WEB DL XviD-FUM [ettv]} has a unsatisfactory video quality according to the comments on the page of the torrent on The Pirate Bay. Generally, an approximately 40-minute TV show file with over 300MB in size can maintain a high level of video quality. It also takes much less time to download than 720p or 1080p HD versions, so most users prefer to download this smaller file. A torrent with larger swarm size will attract more potential peers. And peers in a larger swarm will enjoy better downloading speed. 

\subsection{IP Participation at the Episode Level}

\begin{table}[!h]
\begin{footnotesize}
\renewcommand{\arraystretch}{1.0}
\caption{IP Participation at Episode Level}
\label{episodelevel}
\makebox[\linewidth]{
\begin{tabular}{|c|c|c|}
\hline
\textbf{Episode} & \textbf{Distinct IPs} & \textbf{Overall \%}\\
\hline
Breaking Bad S05E09 & 1,954,961 & 31.03\%\\
\hline
Breaking Bad S05E10 & 1,943,499 & 30.85\%\\
\hline
Dexter S08E07 & 1,280,094 & 20.32\%\\
\hline
Dexter S08E08 & 1,388,402 & 22.04\%\\
\hline
True Blood S06E09 & 1,089,996 & 17.30\%\\
\hline
True Blood S06E10 & 974,839 & 15.47\%\\
\hline
\end{tabular}
}
\end{footnotesize}
\end{table}

For each episode, two torrents were selected for monitoring. Table~\ref{episodelevel} lists the number of distinct active IPs and overall percentage. The episodes from each individual series listed all show a similarity in swarm size. The second selected episode for each torrent (downloaded on 19 August 2013) were only tracked for 24 days - one week less than the first selected ones. As a result, the number of distinct active IPs between the two episodes of Breaking Bad and True Blood may be even closer. Considering the episodes monitored appear late in their respective seasons, a significant impact on swarm size between episodes may be observed if the season or even the TV show has been on the air for a short period of time.

\subsection{Cross-swarm Analysis at the Episode Level}
\begin{figure}
\centering
\begin{tabular}{cc}
\subfloat[Breaking Bad]{\includegraphics[width=4cm]{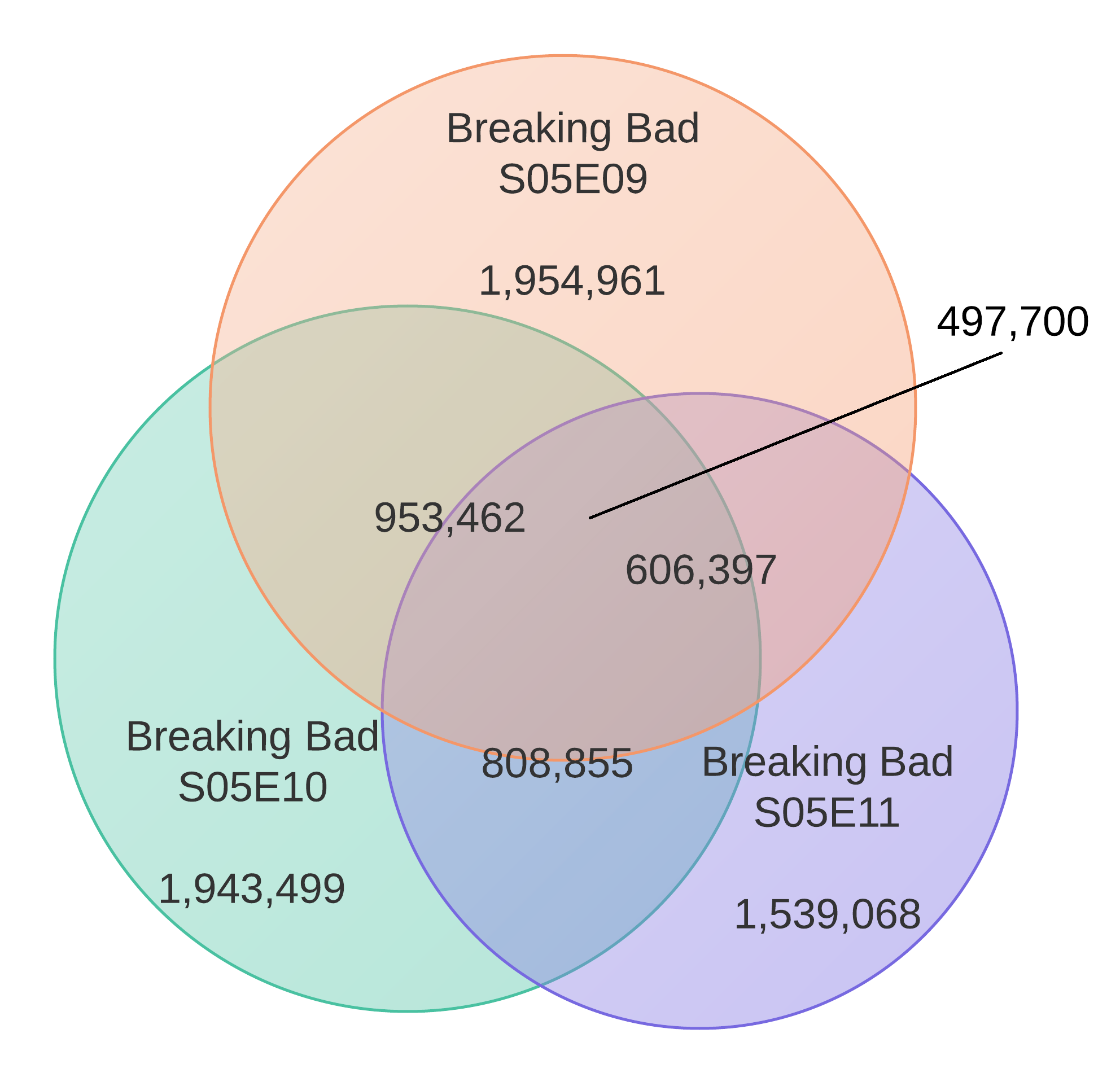}}
\subfloat[Dexter]{\includegraphics[width=4cm]{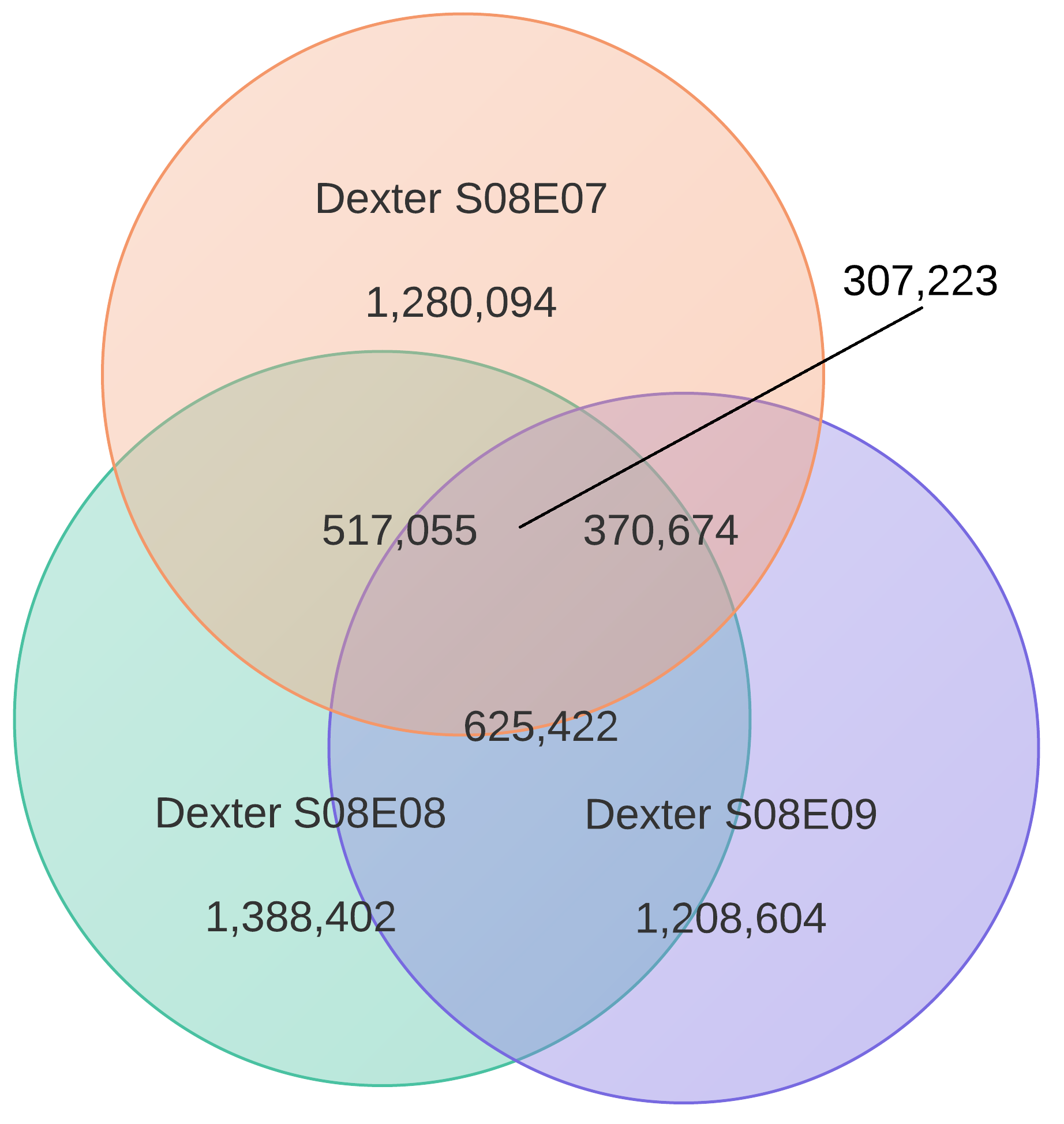}}\\
\subfloat[True Blood]{\includegraphics[width=4cm]{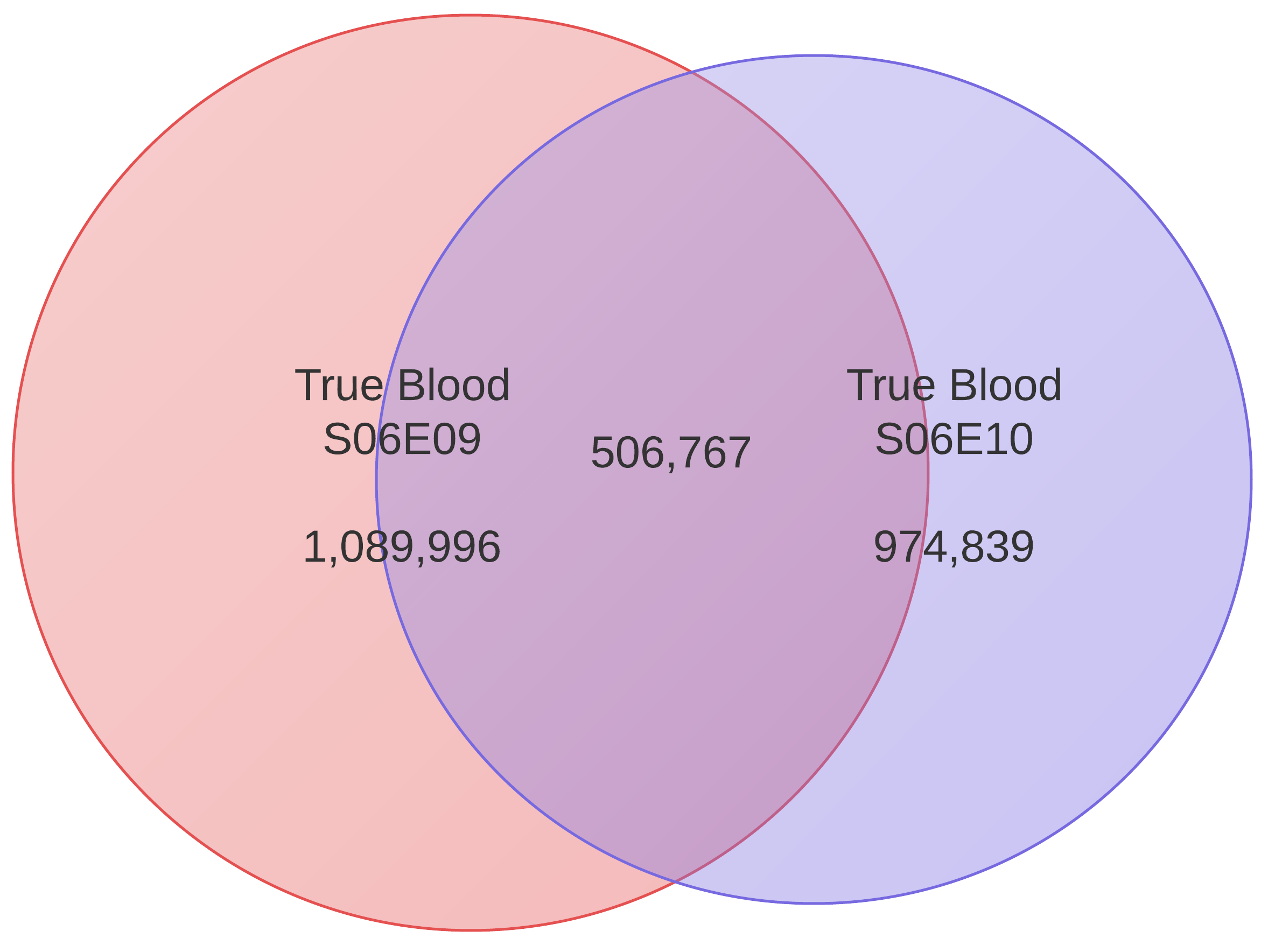}}
\end{tabular}
\caption{Cross-swarm Participation at the Episode Level}
\label{fig_sim_week1_series}
\end{figure}
Cross-swarm participation at episode level is shown in Figure~\ref{fig_sim_week1_series}. Over 40\% of the peers identified in the first episode of each series investigated were also active in the subsequent episode of that series. Among all the peers participating in Breaking Bad and Dexter respectively, over 10\% were active in all three episodes. This rate may be higher if the last two selected episodes of each TV show were also tracked for a entire month. Only a very small proportion (less than 3\%) of the peers participated across the three series.

\subsection{Cross-swarm Analysis at the Series Level}
\begin{figure}[!h]
\centering
\includegraphics[width=5cm]{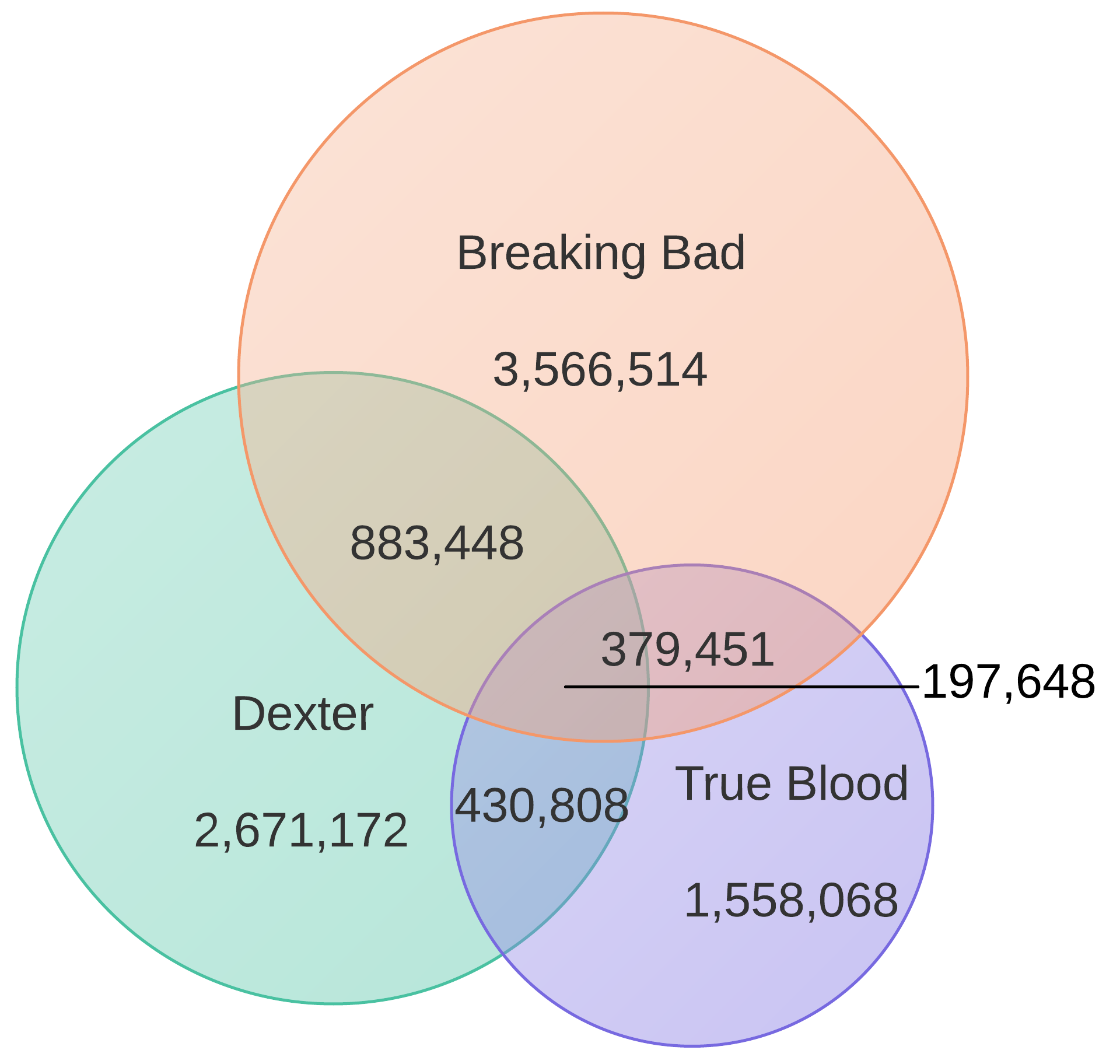}
\caption{Cross-swarm Participation at TV Show Level}
\label{serieslevel}
\end{figure}
Cross-swarm participation at the series level, as shown in Figure~\ref{serieslevel}, indicates that 16.50\% of peers participating in Breaking Bad or Dexter appeared across both shows. The proportion is 11.34\% for Dexter and True Blood and only 8\% for Breaking Bad and True Blood. The similarity in genre between Breaking Bad and Dexter (crime drama, psychological thriller and dark comedy) may partly account for the high cross-swarm participation between the shows. The genre of True Blood is supernatural drama, fantasy and horror.

\begin{figure*}
\centering
\begin{minipage}[b]{0.8\textwidth}

\includegraphics[width=\textwidth]{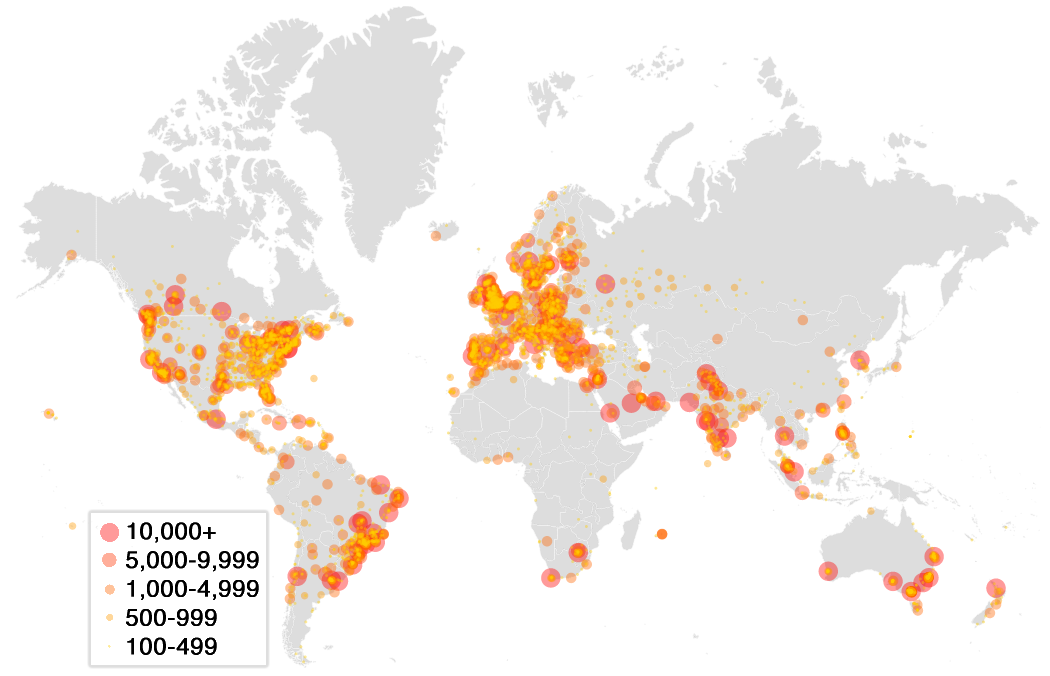}

\caption{Global Map of the IP Distribution at City Level}
\label{globalcitylevel}
\end{minipage}\qquad
\end{figure*}

A global view of IP distribution of the three TV series at a city level is shown in Figure~\ref{globalcitylevel}. Cities with more IPs detected are represented in a darker red colour, while those with smaller numbers are represented in yellow. Cities with less than 100 IPs detected are not portrayed for the purpose of clarity. 
The top 10 cities detected are listed in Table~\ref{tablethreeseries} and generally correspond to places of high population density. The top 10 cities identified may seem to not correlate to the country statistics previously outlined. This is due to the resolution of the IP address geolocation database used. For example, large US cities may appear as multiple different cities in the database due to a higher resolution of IP lookup, e.g., New York city appears under ``New York'' and under each of its components boroughs (Manhattan, The Bronx, Brooklyn, Queens and Staten Island).

\begin{table}[!h]
\begin{footnotesize}
\renewcommand{\arraystretch}{1.0}
\caption{Top 10 Cities for the Three TV Shows}
\label{tablethreeseries}
\makebox[\linewidth]{
\begin{tabular}{|c|c|}
\hline
\textbf{City} & \textbf{Number of distinct IPs}\\
\hline
Athens, Greece & 92,866\\
\hline
London, United Kingdom & 65,203\\
\hline
Perth, Australia & 53,386\\
\hline
Brisbane, Australia & 49,144\\
\hline
Mumbai, India & 48,027\\
\hline
Toronto, Canada & 45,828\\
\hline
Sydney, Australia & 42,899\\
\hline
Islamabad, Pakistan & 41,850\\
\hline
Melbourne, Australia & 38,469\\
\hline
Delhi, India & 38,432\\
\hline
\end{tabular}
}
\end{footnotesize}
\end{table}


\subsubsection{United States}
\begin{figure}[!b]
\centering
\includegraphics[width=0.5\textwidth]{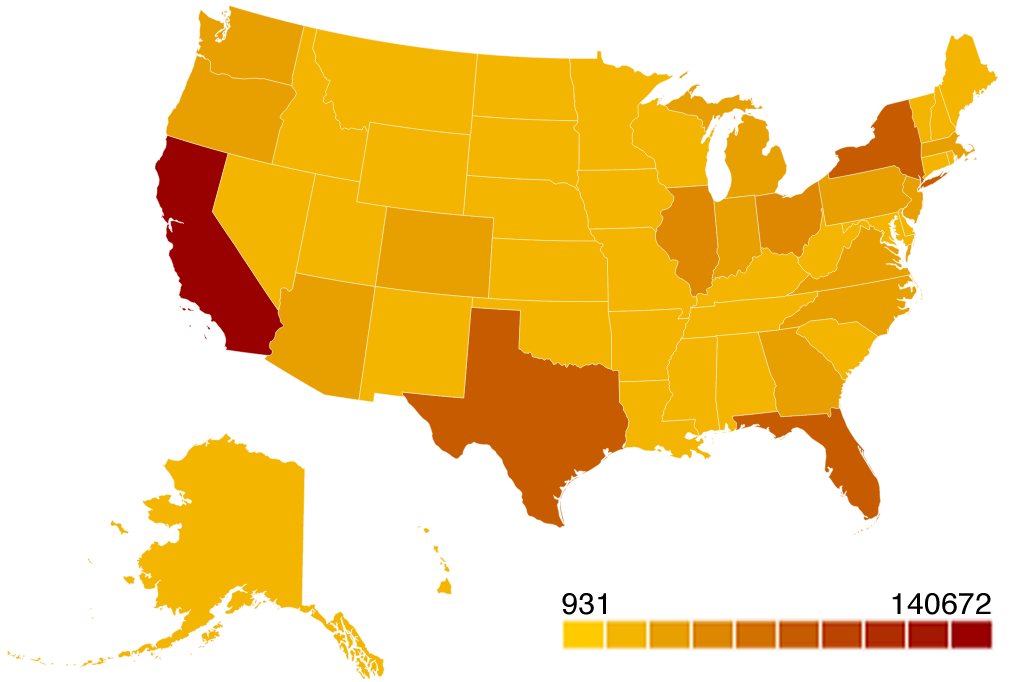}
\caption{US Heatmap of the IP Distribution at State Level}
\label{usheatmap}
\end{figure}

Following population distribution, most of the peers detected from the United States are concentrated on both the east and west coasts. Although the United States accounted for 15.37\% of the total IPs detected, the peers are quite scattered through the country. Los Angeles, New York, Brooklyn, Chicago and San Francisco are the only 5 US cities that had over 10,000 IPs detected. Los Angeles ranks 46th in peer count among the cities globally. At state level, California, New York and Florida are the top 3 states with the most peers detected as shown in Figure~\ref{usheatmap}. They accounted for 14.56\%, 7.99\% and 7.07\% of the detected peers from US respectively.
Although 3,286 US ISPs were detected, the top 5 ISPs cover 57.56\% of the peers in US and the top 20 ISPs cover 82.05\% of the peers. That is to say, a small proportion of ISPs takes most of the traffic burden of BitTorrent file-sharing in US.

%

%

\subsection{Temporal Trend Analysis}

\begin{figure}[!h]
\centering
\includegraphics[width=0.5\textwidth]{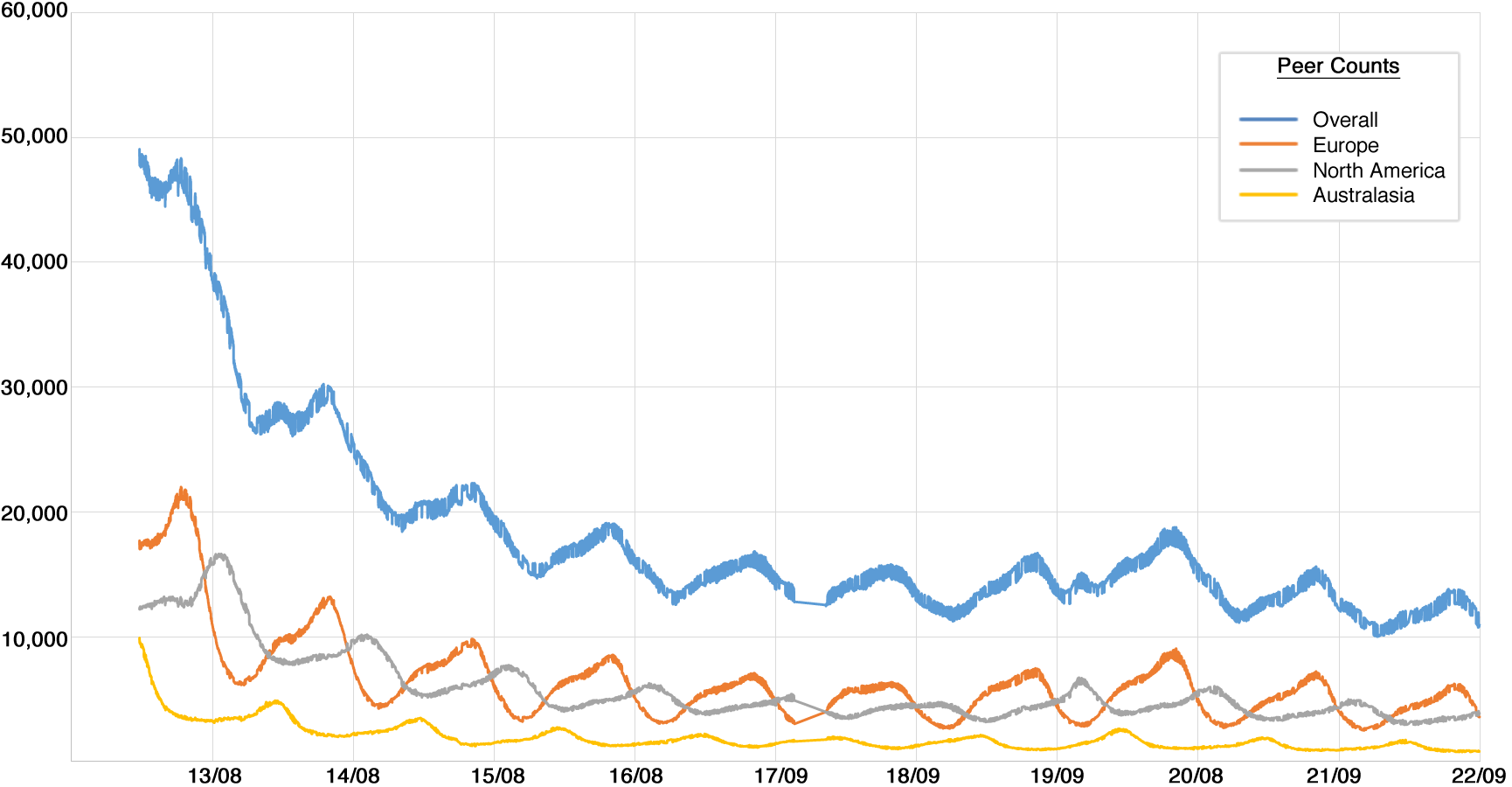}
\caption{The Fluctuation of Swarm Size over a Nine Day Timeframe}
\label{fluctuationsmaller}
\end{figure}

To better analyse the partial fluctuation and the influencing factors, the torrent with the most activity, \texttt{Breaking.Bad.S05E09.HDTV.x264-ASAP.mp4}, was chosen for further analysis, as can be seen in Figure~\ref{fluctuationsmaller}. This graph shows the change of the swarm size from the beginning of the data gathering process to 22 August 2013. In this figure, the size of the US and Canada is consistently smaller than the European size, while the Australian size is the smallest among the three. This may explain why the trend of the swarm size mainly follows that of the European swarm size. 
In comparison, the effect of Australian swarm is nearly negligible. For each region or country analysed, the swarm size from that area rose in the morning and fell at night. The peak generally occurred between 8pm to 9pm (local time). Due to the time difference between the three regions/countries, three distinct peaks were identified daily, as can also be seen in Figure~\ref{fluctuationsmaller}.

On the 19th August 2013 the torrents of the following episode were uploaded on The Pirate Bay, which may well affect the swarm size of the previous episode. The swarm size on 19 August was slightly larger than that on 18 August at every instance measured. This is likely because some peers were reminded to download the previous episode when they saw the latest episode released on The Pirate Bay. 
Another interesting point is that there were unusually two peaks on 19 August. We speculate the smaller peak was produced mainly by the peers from US and Canada because the two peaks occurs almost at the same time and the sum of the three swarm size is very close to the overall total. Time of day (mid-evening) and a boost from the release of the latest episode together contribute to the smaller peak.

\section{Conclusion}
The network analysis outlined as part of this paper aimed to identify the characteristics of IP distribution and measure the temporal trend of the swarm size in BitTorrent network. During one month, 6 torrents of 3 popular TV shows were tracked, with 6,299,695 distinct active IP addresses detected. This paper reveals the characteristics of cross-swarm IP participation, TV show popularity at country, state and city level, as well as fluctuation of swarm size. The findings outlined provide a real and comprehensive analysis on different aspects of sharing of TV show files for later researchers or analysts. 

\subsection{Future Work}
\label{futurework}
Future work can involve a number of expansions in terms of improving the comprehensiveness of the data gathered:
\begin{enumerate}
\item The gathering of data for additional popular content -- A more comprehensive landscape can be modelled through a more large scale data gathering process. The flexibility of the data gathering process allows for current trends across content types to be easily added.
\item Calculate the average downloading time of each peer -- Based on when a peer first appears in the swarm and when he permanently exits the swarm, the average download time and corresponding bandwidth capabilities of the peer can be estimated. Correlating this data alongside the geolocation, time, swarm size and other factors should produce some interesting characteristics of the BitTorrent landscape.
\end{enumerate}

\bibliographystyle{IEEEtran}

\begin{thebibliography}{10}
\providecommand{\url}[1]{#1}
\csname url@samestyle\endcsname
\providecommand{\newblock}{\relax}
\providecommand{\bibinfo}[2]{#2}
\providecommand{\BIBentrySTDinterwordspacing}{\spaceskip=0pt\relax}
\providecommand{\BIBentryALTinterwordstretchfactor}{4}
\providecommand{\BIBentryALTinterwordspacing}{\spaceskip=\fontdimen2\font plus
\BIBentryALTinterwordstretchfactor\fontdimen3\font minus
  \fontdimen4\font\relax}
\providecommand{\BIBforeignlanguage}[2]{{%
\expandafter\ifx\csname l@#1\endcsname\relax
\typeout{** WARNING: IEEEtran.bst: No hyphenation pattern has been}%
\typeout{** loaded for the language `#1'. Using the pattern for}%
\typeout{** the default language instead.}%
\else
\language=\csname l@#1\endcsname
\fi
#2}}
\providecommand{\BIBdecl}{\relax}
\BIBdecl

\bibitem{bitsync}
\BIBentryALTinterwordspacing
{BitTorrent Inc.} (2013) Bittorrent sync user manual. [Online; accessed
  February 2014]. [Online]. Available:
  \url{http://www.bittorrent.com/help/manual/}
\BIBentrySTDinterwordspacing

\bibitem{bitsyncapi}
------. (2013) {{BitTorrent Sync Developer API}}.
  http://www.bittorrent.com/sync/developers/api. [Online; accessed February
  2014].

\bibitem{bitsyncstats}
\BIBentryALTinterwordspacing
------. (2013) {{BitTorrent Sync Article}}. [Online; accessed February 2014].
  [Online]. Available: \url{http://blog.bittorrent.com/2013/12/05/
  bittorrent-sync-hits-2-million-user-mark/}
\BIBentrySTDinterwordspacing

\bibitem{cohen2008bittorrent}
\BIBentryALTinterwordspacing
B.~Cohen. (2008) The bittorrent protocol specification. [Online; accessed
  February 2014]. [Online]. Available:
  \url{http://bittorrent.org/beps/bep_0003.html/}
\BIBentrySTDinterwordspacing

\bibitem{cohen2003incentives}
------, ``Incentives build robustness in bittorrent,'' in \emph{Proceedings of
  the Workshop on Economics of Peer-to-Peer systems}, vol.~6, 2003, pp. 68--72.

\bibitem{sit2002security}
E.~Sit and R.~Morris, ``Security considerations for peer-to-peer distributed
  hash tables,'' in \emph{Peer-to-Peer Systems}.\hskip 1em plus 0.5em minus
  0.4em\relax Springer, 2002, pp. 261--269.

\bibitem{farina2014}
J.~Farina, M.~Scanlon, and M.-T. Kechadi, ``Bittorrent sync: First impressions
  and forensic implications,'' in \emph{Digital Forensic Research Workshop EU
  (DFRWS EU 2014) [Pending Publication]}, May 2014.

\bibitem{layton2010investigation}
R.~Layton and P.~Watters, ``Investigation into the extent of infringing content
  on bittorrent networks,'' \emph{Internet Commerce Security Laboratory}, 2010.

\bibitem{scanlon2010week}
M.~Scanlon, A.~Hannaway, and M.-T. Kechadi, ``A week in the life of the most
  popular bittorrent swarms,'' \emph{5th Annual Symposium on Information
  Assurance (ASIA'10)}, 2010.

\bibitem{le2010spying}
S.~Le~Blond, A.~Legout, F.~Lefessant, W.~Dabbous, and M.~A. Kaafar, ``Spying
  the world from your laptop: identifying and profiling content providers and
  big downloaders in bittorrent,'' in \emph{Proceedings of the 3rd USENIX
  conference on Large-scale exploits and emergent threats: botnets, spyware,
  worms, and more}.\hskip 1em plus 0.5em minus 0.4em\relax USENIX Association,
  2010, pp. 4--4.

\bibitem{Chung201281}
H.~Chung, J.~Park, S.~Lee, and C.~Kang, ``Digital forensic investigation of
  cloud storage services,'' \emph{Digital Investigation}, vol.~9, no.~2, pp. 81
  -- 95, 2012.

\bibitem{Martini2013287}
B.~Martini and K.-K.~R. Choo, ``Cloud storage forensics: owncloud as a case
  study,'' \emph{Digital Investigation}, vol.~10, no.~4, pp. 287 -- 299, 2013.

\bibitem{reddit}
Reddit. (2014) Btsecrets. http://www.reddit.com/r/btsecrets. [Online; accessed
  February 2014].

\bibitem{maxmind}
\BIBentryALTinterwordspacing
M.~Inc. (2014, Jul.) Geolite country database. [Online]. Available:
  \url{http://www.maxmind.com}
\BIBentrySTDinterwordspacing

\bibitem{herrera2007modeling}
O.~Herrera and T.~Znati, ``Modeling churn in p2p networks,'' in
  \emph{Simulation Symposium, 2007. ANSS'07. 40th Annual}.\hskip 1em plus 0.5em
  minus 0.4em\relax IEEE, 2007, pp. 33--40.

\end{thebibliography}


\begin{thebibliography}{1}
\bibitem{cisco}
Cisco Systems, Inc. ``Cisco Visual Networking Index: Forecast and Methodology, 2012-2017.'' 2013. Retrieved iMay 2014, \url{http://www.cisco.com/en/US/solutions/collateral/ns341/ns525/ns537/ns705/ns827/white_paper_c11-481360.pdf}
\bibitem{paloalto}
Palo Alto Networks. The Application Usage and Threat Report. 2013. Retrieved August 2013, \url{http://researchcenter.paloaltonetworks.com/app-usage-risk-report-visualization}
\bibitem{bt}
BitTorrent, Inc. ``BitTorrent and $\mu$Torrent Software Surpass 150 Million User Milestone.'' 2012. Retrieved May 2014, \url{http://www.bittorrent.com/intl/es/company/about/ces_2012_150m_users}
\bibitem{btspec}
The BitTorrent Protocol Specification. 2008. Retrieved May 2014,\\ \url{http://www.bittorrent.org/beps/bep_0003.html}
\bibitem{tpb}
The Pirate Bay. Retrieved August 2013, \url{http://thepiratebay.se/}
\bibitem{torrentfreak}
TorrentFreak. ``Top 10 Most Pirated TV-Shows of the Season.'' Retrieved May 2014, \url{http://torrentfreak.com/top-10-most-pirated-tv-shows-of-the-season-130622/}
\bibitem{erman}
Erman, David. ``BitTorrent traffic measurements and models.'' Blekinge Institute of Technology, 2005. 18-19, 24
\bibitem{scanlon}
Scanlon, Mark, Alan Hannaway, and Mohand-Tahar Kechadi. ``A week in the life of the most popular BitTorrent swarms.'' In Proceedings of the 5th Annual Symposium on Information Assurance (ASIA 2010), pp. 32-36. 2010.
\bibitem{maxmind}
Maxmind GeoIP2 Geolocational Databases. Retrieved May 2014,\\ \url{http://dev.maxmind.com/geoip/geoip2/geolite2/}
\bibitem{chen}
Chen, Xi, Kai Lin, Biao Wang, and Zhe Yang. ``Active measurements on BitTorrent and eMule ecosystems over the Internet.'' In Consumer Electronics, Communications and Networks (CECNet), 2012 2nd International Conference on, pp. 126-129. IEEE, 2012.
\bibitem{salvador}
Salvador, Paulo, and António Nogueira. ``Study on geographical distribution and availability of bittorrent peers sharing video files.'' In Consumer Electronics, 2008. ISCE 2008. IEEE International Symposium on, pp. 1-4. IEEE, 2008.
\bibitem{zhang}
Zhang, Boxun, Alexandru Iosup, Johan Pouwelse, and Dick Epema. ``Identifying, analyzing, and modeling flashcrowds in bittorrent.'' In Peer-to-Peer Computing (P2P), 2011 IEEE International Conference on, pp. 240-249. IEEE, 2011.
\bibitem{pouwelse}
Pouwelse, Johan, Paweł Garbacki, Dick Epema, and Henk Sips. ``The BitTorrent P2P file-sharing system: Measurements and analysis.'' In Peer-to-Peer Systems IV, pp. 205-216. Springer Berlin Heidelberg, 2005.
\bibitem{guo} 
Guo, Lei, Songqing Chen, Zhen Xiao, Enhua Tan, Xiaoning Ding, and Xiaodong Zhang. ``A performance study of BitTorrent-like peer-to-peer systems.'' Selected Areas in Communications, IEEE Journal on 25, no. 1 (2007): 155-169.
\end{thebibliography}
%



\end{document}